\documentclass{aa}
\usepackage{allrunes}
\usepackage{graphicx}
\usepackage{txfonts}
\usepackage{wasysym}
\usepackage{array}
\usepackage[pdftitle={Transit Least-Squares Survey -- I. -- Discovery and validation of an Earth-sized planet in the four-planet system K2-32 near the 1:2:5:7 resonance}, colorlinks = true, breaklinks = true, citecolor = blue, linkcolor = blue, urlcolor = blue, pdfauthor = {Ren\'{e} Heller}]{hyperref}
\usepackage{esvect}  

\newcolumntype{C}[1]{>{\centering\let\newline\\\arraybackslash\hspace{0pt}}m{#1}}

\begin{document}

\title{Transit least-squares survey}
\subtitle{I. Discovery and validation of an Earth-sized planet in the four-planet system K2-32 near the 1:2:5:7 resonance}
\titlerunning{The Transit Least-Squares Survey -- I. Discovery and validation of K2-32\,e}
\author{Ren\'{e} Heller\inst{1}
\and
Kai Rodenbeck\inst{1,2}
\and
Michael Hippke\inst{3}
          }

   \institute{Max Planck Institute for Solar System Research, Justus-von-Liebig-Weg 3, 37077 G\"ottingen, Germany, \href{mailto:heller@mps.mpg.de}{heller@mps.mpg.de}
   \and
   \scalebox{.983}[1.0]{Inst. for Astrophysics, Georg-August-Univ. G\"ottingen, Friedrich-Hund-Platz 1, 37077 G\"ottingen, Germany, \href{mailto:rodenbeck@mps.mpg.de}{rodenbeck@mps.mpg.de}}
   \and   
   Sonneberg Observatory, Sternwartestra{\ss}e 32, 96515 Sonneberg, Germany, \href{mailto:michael@hippke.org}{michael@hippke.org}
             }

   \date{Received 14 February 2019 / Accepted 30 March 2019}


\abstract{
We apply for the first time the Transit Least-Squares ({\tt TLS}) algorithm to search for new transiting exoplanets. {\tt TLS} has been developed as a successor to the Box Least-Squares ({\tt BLS}) algorithm, which has served as a standard tool for the detection of periodic transits. In this proof-of-concept paper, we demonstrate how {\tt TLS} finds small planets that have previously been missed. We showcase {\tt TLS}'s capabilities using the K2 {\tt EVEREST}-detrended light curve of the star K2-32 (EPIC\,205071984), which has been known to have three transiting planets. {\tt TLS} detects these known Neptune-sized planets K2-32\,b, d, and c in an iterative search and finds an additional transit signal with a high signal detection efficiency (SDE$_{\rm TLS}$) of 26.1 at a period of $4.34882_{-0.00075}^{+0.00069}$\,d. We show that this additional signal remains detectable (SDE$_{\rm TLS}$~=~13.2) with {\tt TLS} in the {\tt K2SFF} light curve of K2-32, which includes a less optimal detrending of the systematic trends. The signal is below common detection thresholds, however, if searched with {\tt BLS} in the {\tt K2SFF} light curve (SDE$_{\rm BLS}$~=~8.9) as in previous searches. Markov Chain Monte Carlo sampling with the {\tt emcee} software shows that the radius of this candidate is ${1.01}_{-0.09}^{+0.10}\,R_\oplus$. We analyze its phase-folded transit light curve using the {\tt vespa} software and calculate a false positive probability ${\rm FPP}=3.1~\times~10^{-3}$. Taking into account the multiplicity boost of the system, we estimate an ${\rm FPP}<3.1~\times~10^{-4}$, which formally validates K2-32\,e as a planet. K2-32 now hosts at least four planets that are very close to a 1:2:5:7 mean motion resonance chain. The offset of the orbital periods of K2-32\,e and b from a 1:2 mean motion resonance is in very good agreement with the sample of transiting multiplanet systems from {\it Kepler}, lending further credence to the planetary nature of K2-32\,e. We expect that {\tt TLS} can find many more transits of Earth-sized and even smaller planets in the {\it Kepler} and {\it K2} data that have hitherto remained undetected with algorithms that search for box-like signals.}


\maketitle


\section{Introduction}
\label{sec:introduction}

\begin{figure*}[t]
\centering
\includegraphics[width=1.01\linewidth]{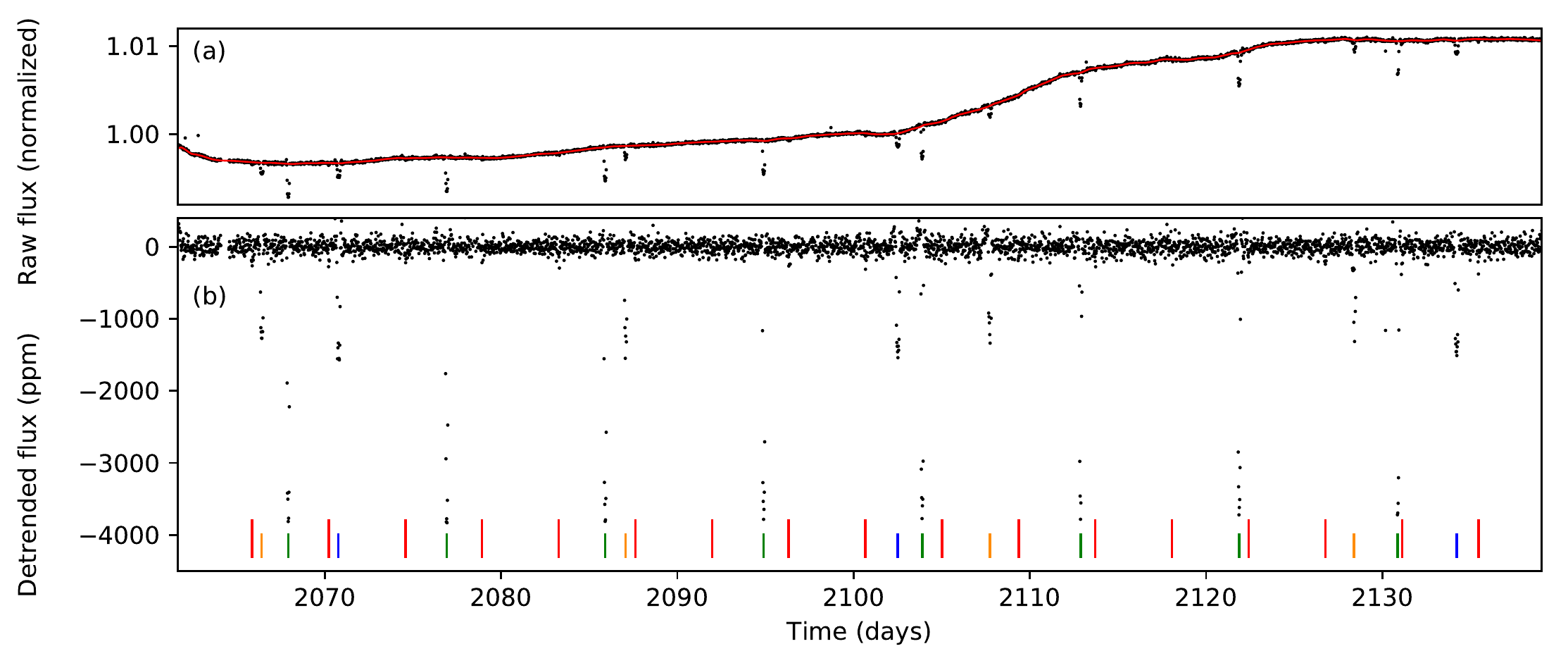}
\caption{{\it K2} light curve of K2-32. \textbf{(a)} After correction for systematic effects with {\tt EVEREST}. The red line shows our running median filter. \textbf{(b)} Detrended light curve obtained by dividing the {\tt EVEREST} light curve by the running median. Transits detected with {\tt TLS} are highlighted with green (K2-32\,b), magenta (K2-32\,c), blue (K2-32\,d), and stretched red (K2-32\,e) vertical bars.}
\label{fig:data_reduction}
\end{figure*}

The data from the {\it Kepler} primary mission \citep[{\it K1};][]{2010Sci...327..977B}, which operated from 2009 to 2013, and from the repurposed {\it K2} mission \citep{2014PASP..126..398H}, which worked from 2014 to 2018, have both been subject to extensive transit searches. Most of their confirmed or validated planets (2338 from {\it K1} and 359 from K2)\footnote{\href{https://exoplanetarchive.ipac.caltech.edu/docs/counts_detail.html}{https://exoplanetarchive.ipac.caltech.edu/docs/counts\_detail.html}\\on 28 March 2019}\,and of the candidates that are yet to be confirmed (2423 from {\it K1} and 536 from {\it K2}) have been found using the Box Least-Squares ({\tt BLS}) transit search algorithm \citep{2002A&A...391..369K} or similar algorithms searching for box-like flux decreases in stellar light curves \citep{2013ApJS..204...24B,2016ApJS..222...14V,2016ApJS..226....7C,2018ApJS..239....5C,2018AJ....155..136M,2018AJ....156...78L,2018AJ....156..277L,2018AJ....156...22Y,2018MNRAS.474.4603V}.

We developed the Transit Least-Squares ({\tt TLS}) algorithm \citep{2019A&A...623A..39H} as the successor of {\tt BLS} in order to be even more sensitive to smaller possibly sub-Earth-sized planets. Instead of searching for box-like flux decreases in the light curve, {\tt TLS} is based on an analytical transit model with stellar limb darkening \citep{1977A&A....61..809M,2002ApJ...580L.171M}. The signal detection efficiency of {\tt TLS} is consequently improved compared to {\tt BLS}, while the false-positive rate is also suppressed \citep{2019A&A...623A..39H}. Here we use {\tt TLS} to search for so far unknown planets in the {\it K2} data of K2-32 (EPIC\,205071984), and we present our first discovery from our new data analysis campaign, the {\tt TLS} Survey.

 Three planets have previously been reported around K2-32 by \citet{2016ApJS..222...14V}, who formally designated them as candidates, and by \citet{2016ApJS..226....7C}, who validated them as planets K2-32\,b, c, and d using additional high-resolution spectroscopy and independent stellar photometry to feed the statistical vetting software {\tt vespa} \citep{2012ApJ...761....6M,2015ascl.soft03011M}. \citet{2018AJ....155..136M}, also using {\tt vespa} and additional adaptive optics observations, again detected the three transiting objects and determined their probabilities of being an eclipsing binary, background eclipsing binary, or hierarchical eclipsing binary each to be $<10^{-4}$. Most important, all these previous detections of K2-32\,b, c, and d were achieved in searches for box-like transit signals. \citet{2016ApJS..222...14V} used {\tt BLS}, \citet{2016ApJS..226....7C} used the {\tt TERRA} software \citep{2013PNAS..11019273P}, which includes a search for box-like transit signals very much like {\tt BLS}, and \citet{2018AJ....155..136M} also used {\tt BLS}. K2-32\,b, c, and d have also been confirmed through stellar radial velocity measurements by \citet{2016ApJ...823..115D} and \citet{2017AJ....153..142P}.

\begin{figure}
\centering
\includegraphics[width=0.85\linewidth]{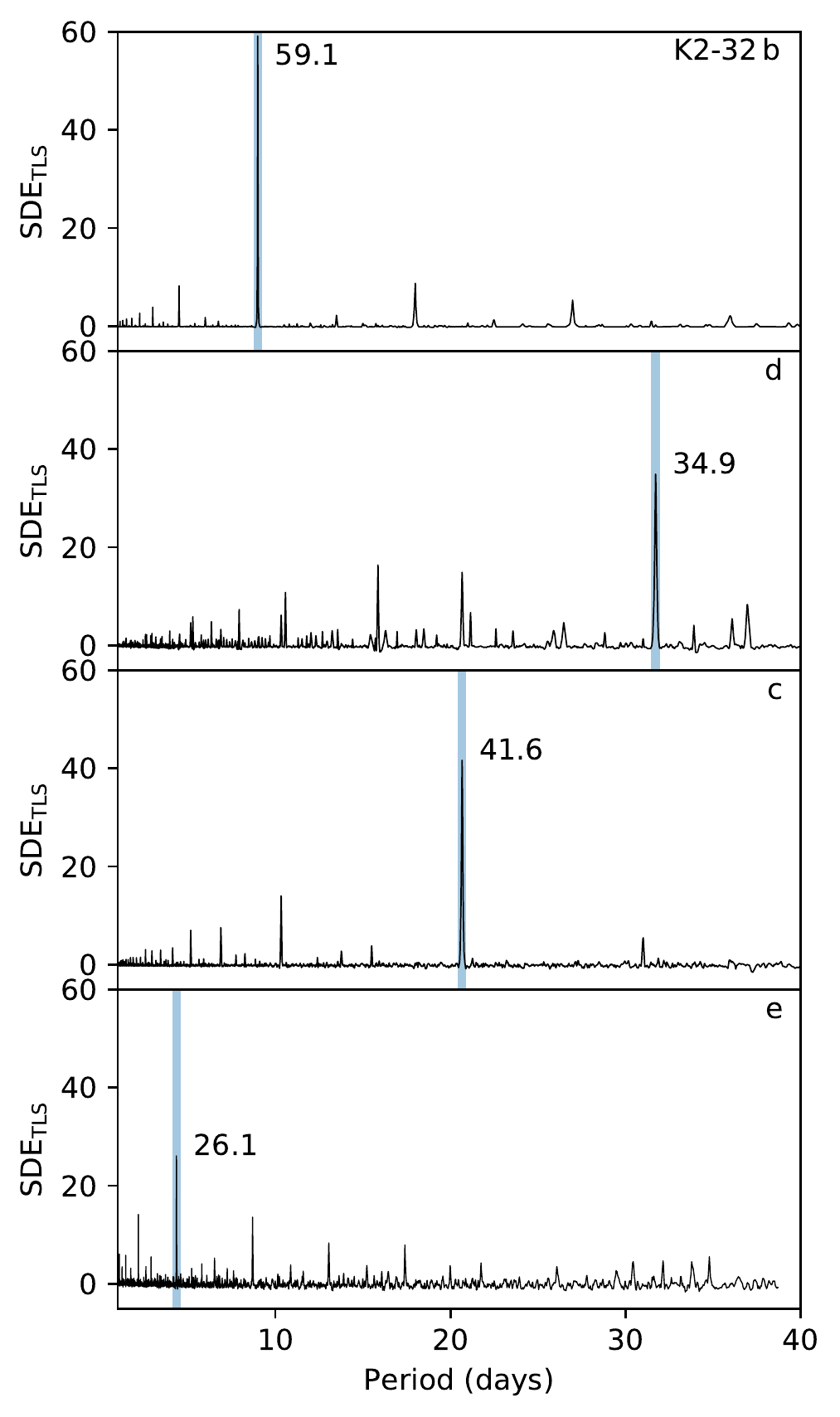}
\caption{Signal detection efficiencies of our successive transit search after masking out previously detected transits.}
\label{fig:SDE}
\end{figure}

\begin{figure}[h]
\centering
\includegraphics[width=\linewidth]{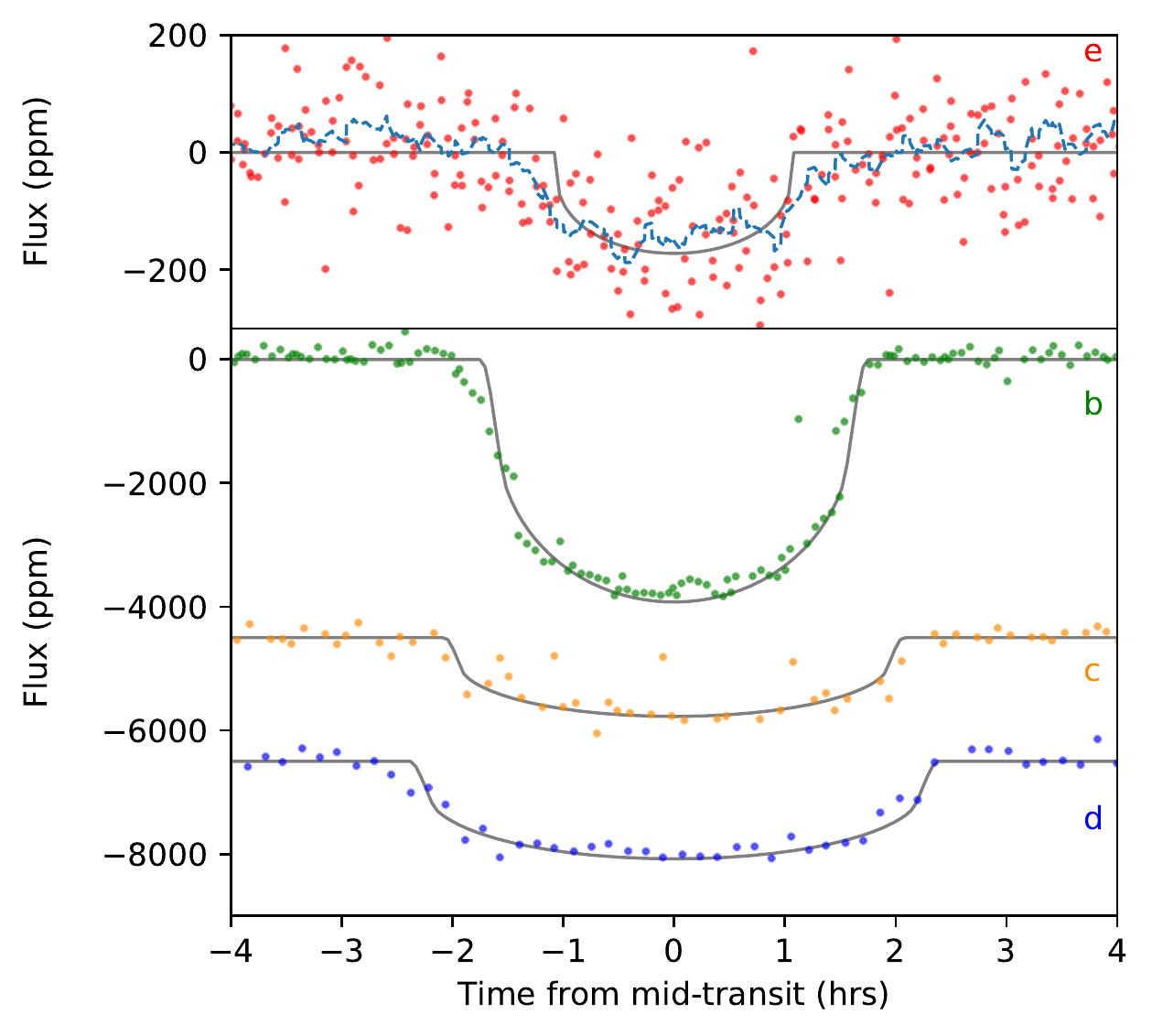}
\caption{Phase-folded transit light curves of all planets. Dots show the {\it K2} data, lines represent our best-fit MCMC models. The ordinates in the top and bottom panels have different scales. In the top panel, the transit dip of K2-32\,e is shown together with a sliding mean (dashed line) of 11 cadences in width.}
\label{fig:fold}
\end{figure}

\begin{figure}[h]
\centering
\includegraphics[width=1.\linewidth]{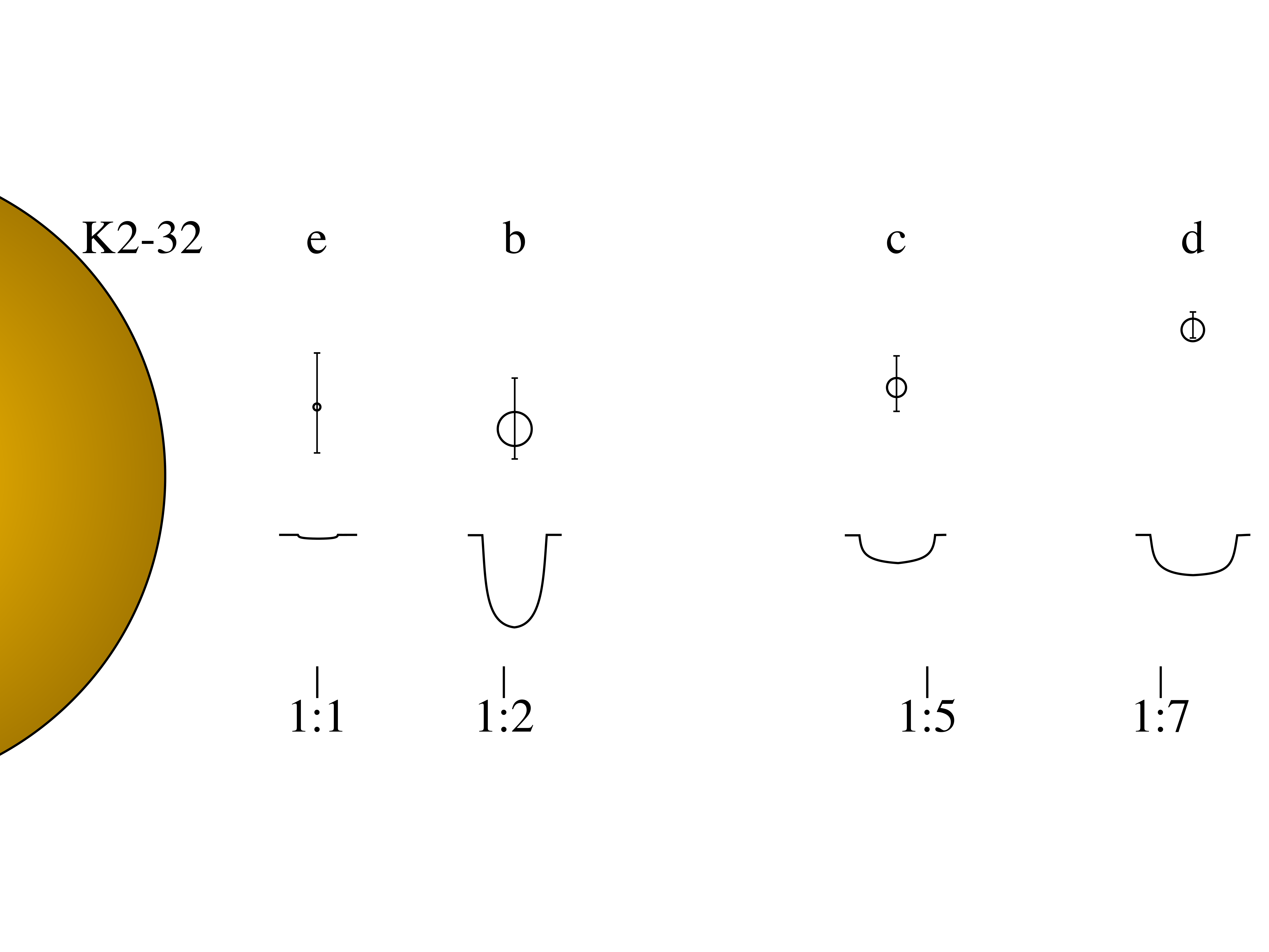}
\caption{System architecture. Stellar and planetary radii are to scale. Planetary distances to the star are mutually to scale, but not with respect to the radii. The shapes of the transit light curves are to scale as well. Orbital resonances are indicated with respect to the innermost planet K2-32\,e. The error bars denote our uncertainties in the transit impact parameter (Table~\ref{tab:MCMC}).}
\label{fig:architecture}
\end{figure}

\begin{figure*}
\centering
\includegraphics[width=.42\linewidth]{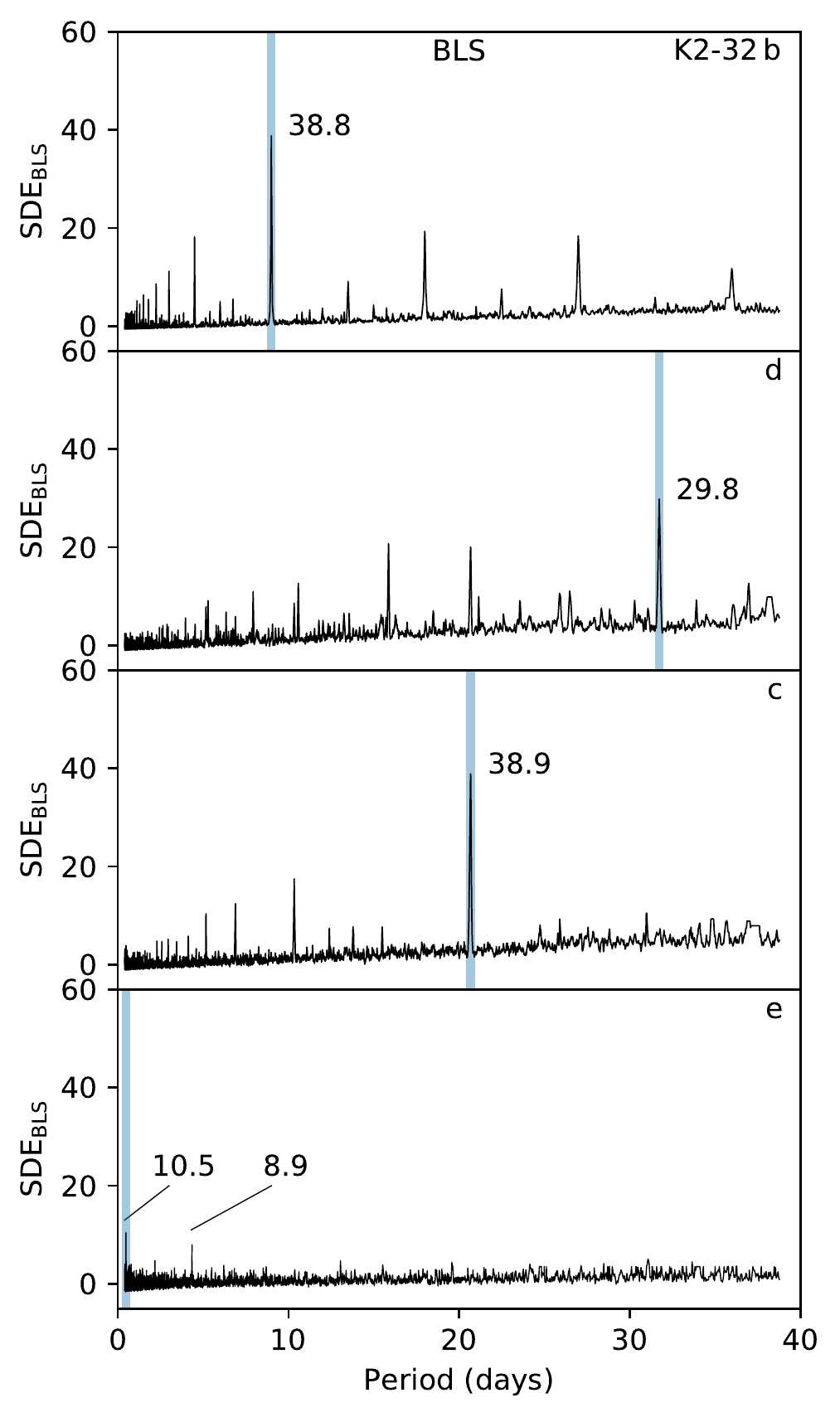}
\hspace{0.8cm}
\includegraphics[width=.42\linewidth]{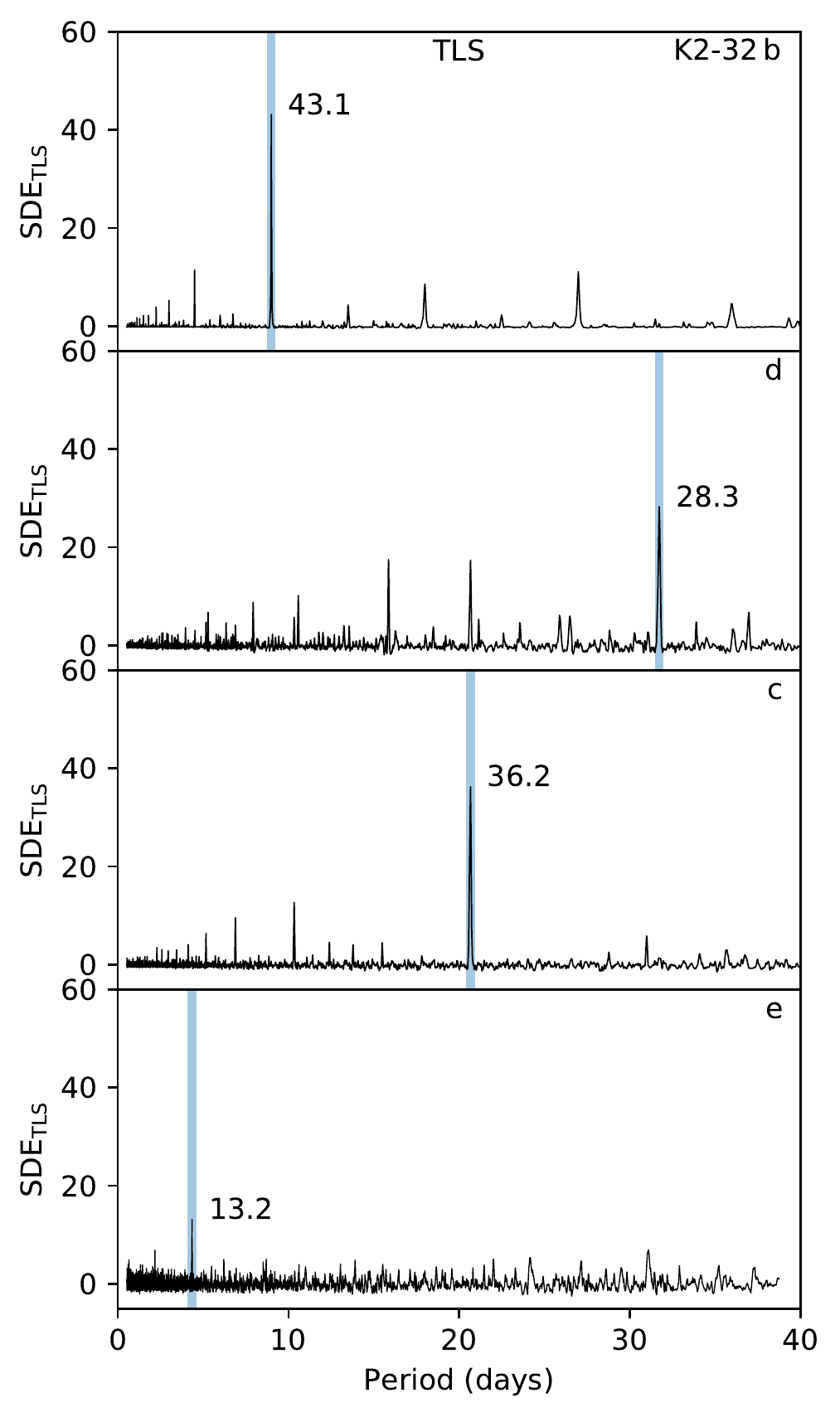}
\caption{Iterative transit search using {\tt K2SFF} data with {\tt BLS} (left) and {\tt TLS} (right). Planets K2-32\,b, d, and c are successively detected with strong signals (see labels) with both {\tt BLS} and {\tt TLS}. Using a common detection threshold value of 9, however, planet e is not detected with {\tt BLS} (SDE$_{\rm BLS}$~=~8.9), but it is detected with {\tt TLS} (SDE$_{\rm TLS}$~=~13.2). The major peak near $P=0.4$\,d with (SDE$_{\rm BLS}$~=~10.5) in the fourth iteration of {\tt BLS} is a false positive (blue marker). It overshadows the true transit signal near $P=4.3\,$d with (SDE$_{\rm BLS}$~=~8.9).}
\label{fig:k2sff}
\end{figure*}

\section{Methods}
\label{sec:methods}

\subsection{Target selection}

In this paper, we move on from the testing phase of {\tt TLS} \citep{2019A&A...623A..39H} and apply {\tt TLS} to real light curves. In this initial phase of the {\tt TLS} Survey, we restrict ourselves to multiplanet systems from the {\it Kepler} mission because they have been shown to exhibit extremely low false-positive probabilities (FPP) \citep{2012ApJ...750..112L}. Among the systems that we studied, K2-32 with its three previously known planets stood out with our detection of a highly significant fourth transit signal, the method of which we describe in the following.

\subsection{Transit search}

The light curve of K2-32 from campaign 2 contains 77.4\,d of almost uninterrupted observations with 3527 useful exposures at a cadence of 30\,min. We ignored the first 64 exposures in the light curve because they are affected by strong systematics. We removed outliers, defined as data points $>3\,\sigma$ above the running mean, from the publicly available\footnote{\href{https://archive.stsci.edu/hlsps/everest/v2/c02/205000000/71984/hlsp_everest_k2_llc_205071984-c02_kepler_v2.0_lc.fits}{https://archive.stsci.edu/hlsps/everest/v2/c02/205000000/71984/\\hlsp\_everest\_k2\_llc\_205071984-c02\_kepler\_v2.0\_lc.fits}} {\it K2} light curve of K2-32, which has been corrected for instrumental effects with {\tt EVEREST} \citep{2016AJ....152..100L,2018AJ....156...99L} (Fig.~\ref{fig:data_reduction}a). We then removed stellar variability and other trends using a median filter with a window size of 1\,d. The resulting detrended light curve is shown in Fig.~\ref{fig:data_reduction}(b).

We tested other window sizes and found that a width of 1\,d offers the best compromise between both sufficient removal of unwanted variability and preservation of transit signals with durations of up to several hours. Kepler's third law of motion \citep{1619ikhm.book.....K} predicts that planets with orbital periods $<80\,$\,d around sun-like stars have transit durations shorter than about 8\,hr \citep[see Fig.~5 in][]{2019A&A...623A..39H}. For K2-32 in particular, the maximum duration of a planet with a single transit in the 80\,d K2 light curve on a circular orbit is 5.7\,hr, and for a planet to exhibit two transits (hence $P<40$\,d) the maximum transit duration is 4.8\,hr, suggesting that a 1\,d width for our median filter hardly affects physically plausible transit signals.

We applied the publicly available\footnote{\href{https://github.com/hippke/tls}{https://github.com/hippke/tls}} {\tt python} implementation of {\tt TLS} \citep{2019A&A...623A..39H} using the stellar-limb darkening, mass, and radius estimates available from the EPIC catalog \citep{2016ApJS..224....2H}. We used {\tt TLS} version 1.0.16 in its default parameterization, but set the maximum trial period equal to the length of the light curve in order to search for possible single transits.

\subsection{Markov chain Monte Carlo analysis of the transits}

To refine the planetary parameters we used the Markov chain Monte Carlo (MCMC) sampler {\tt emcee} \citep{2013PASP..125..306F}. As inputs to {\tt emcee}, we provided the times of the mid-points of the first transit ($T_0$) and the orbital period ($P$) obtained with {\tt TLS} for each planet. $P$, $T_0$, the planet-to-star radius ratio ($R_{\rm p}/R_{\rm s}$), and the transit impact parameter ($b$) served as model parameters for each planet, while the stellar density ($\rho_{\rm s}$) and two limb-darkening coefficients for a quadratic limb darkening law \citep{2013MNRAS.435.2152K} were global parameters for all transits. We ran the MCMC analysis with 100 walkers with each walker performing $200,000$ steps. The first half of each walk was discarded to ensure that we preserved only burned-in MCMC chains.

\section{Results}
\label{sec:results}

\begin{table*}
\caption{Characterization of the new planet K2-32\,e from MCMC model fitting to the full set of transits in the {\tt EVEREST} light curve of K2-32.}
\def\arraystretch{1.4}
\label{tab:MCMC}
\centering
\begin{tabular}{c | p{0.9cm}p{2.04cm} | p{1.05cm}p{2.31cm} | p{1.1cm}p{2.30cm} | p{1.11cm}p{2.25cm}}
\hline\hline
 & \multicolumn{2}{c}{Planet e}  & \multicolumn{2}{c}{Planet b} & \multicolumn{2}{c}{Planet c} & \multicolumn{2}{c}{Planet d}\\
 & \multicolumn{1}{c}{ML\tablefootmark{(a)}}  & \multicolumn{1}{c}{median$_{-{\rm err.}}^{+{\rm err.}}$} & \multicolumn{1}{c}{ML\tablefootmark{(a)}} & \multicolumn{1}{c}{median$_{-{\rm err.}}^{+{\rm err.}}$}  & \multicolumn{1}{c}{ML\tablefootmark{(a)}} & \multicolumn{1}{c}{median$_{-{\rm err.}}^{+{\rm err.}}$}  & \multicolumn{1}{c}{ML\tablefootmark{(a)}} & \multicolumn{1}{c}{median$_{-{\rm err.}}^{+{\rm err.}}$}  \\
\hline
 $P$\,[d] & 4.34911 & $4.34882_{-0.00075}^{+0.00069}$ & 8.991866 & $8.991828_{-0.000084}^{+0.000083}$ & 20.66157 & $20.66186_{-0.00098}^{+0.00102}$ & 31.7145 & $31.7142_{-0.0010}^{+0.0011}$ \\
$T_0$ [d]\tablefootmark{(b)} & $0.8799$ & $0.8860_{-0.0079}^{+0.0085}$ & $2.92698$ & $2.92713_{-0.00034}^{+0.00035}$ & $1.4231$ & $1.4227_{-0.0021}^{+0.0021}$& $5.7923$ & $5.7913_{-0.0017}^{+0.0014}$\\
$R_{\rm p}/R_{\rm s}$ & 0.01080 & $0.01090_{-0.00051}^{+0.00052}$ & 0.05347 & $0.05382_{-0.00068}^{+0.00076}$ & 0.029430 & $0.029664_{-0.00053}^{+0.00058}$ & 0.03534 &	$0.03551_{-0.00062}^{+0.00064}$   \\
$b$ & 0.029 & $0.22_{-0.15}^{+0.17}$ & 0.047 & $0.15_{-0.10}^{+0.16}$ & 0.24 & $0.28_{-0.08}^{+0.10}$ & 0.451 & $0.463_{-0.030}^{+0.057}$ \\
\hline
\end{tabular}
\tablefoot{
           \tablefoottext{a}{ML = value with maximum likelihood.} \tablefoottext{b}{$T_0~=~{\rm BKJD}-2065$\,d with the Barycentric Kepler Julian Day BKJD = BJD - 2,454,833.0\,d.}
          }
\end{table*}

\begin{table*}
\caption{System parameterization of K2-32 and its planets and comparison to literature values.}
\def\arraystretch{1.4}
\label{tab:system}
\centering
\begin{tabular}{c | l l l l l l}
\hline\hline
 & Star & Planet e & Planet b & Planet c & Planet d & Ref.\\
\hline
 $P$\,[d] & & & $8.99213$ & $20.6602$ & $31.7154$ & C16 \\
 & & & $8.99218_{-0.00020}^{+0.00020}$ & $20.65614_{-0.00598}^{+0.00598}$ & $31.71922_{-0.00236}^{+0.00236}$ & D16 \\
 & & & $8.99213_{-0.00016}^{+0.00016}$ & $20.6602_{-0.0017}^{+0.0017}$ & $31.7154_{-0.0022}^{+0.0022}$ & P17 \\
 & & & $8.99194_{-0.00016}^{+0.00016}$ & $20.6616_{-0.0018}^{+0.0017}$ & $31.7151_{-0.0026}^{+0.0022}$ & M18 \\
 & & $4.34882_{-0.00075}^{+0.00069}$ & $8.991828_{-0.000084}^{+0.000083}$ & $20.66186_{-0.00098}^{+0.00102}$ & $31.7142_{-0.0010}^{+0.0011}$ & H19 \\
\hline
$R$& & & $5.62\,R_\oplus$ & $3.32\,R_\oplus$ & $3.77\,R_\oplus$ & C16\tablefootmark{(a)} \\
& $0.87_{-0.05}^{+0.05}\,R_\odot$ & & $5.38_{-0.35}^{+0.35}\,R_\oplus$ & $3.48_{-0.42}^{+0.97}\,R_\oplus$ & $3.75_{-0.40}^{+0.40}\,R_\oplus$ & D16 \\
& $0.845_{-0.035}^{+0.044}\,R_\odot$ & & $5.13_{-0.28}^{+0.28}\,R_\oplus$ & $3.01_{-0.25}^{+0.25}\,R_\oplus$ & $3.43_{-0.35}^{+0.35}\,R_\oplus$  & P17 \\
& $0.839_{-0.026}^{+0.021}\,R_\odot$ & & $5.17_{-0.20}^{+0.16}\,R_\oplus$ & $3.12_{-0.18}^{+0.12}\,R_\oplus$ & $3.41_{-0.26}^{+0.14}\,R_\oplus$  & M18 \\
 & & $1.01_{-0.09}^{+0.10}\,R_\oplus$ & $4.96_{-0.27}^{+0.33}\,R_\oplus$ & $2.74_{-0.16}^{+0.20}\,R_\oplus$ & $3.27_{-0.19}^{+0.23}\,R_\oplus$ & H19\tablefootmark{(b)}  \\
\hline
$M$ & $0.87_{-0.04}^{+0.04}\,M_\odot$ & & $21.1_{-5.9}^{+5.9}\,M_\oplus$ & $<8.1\,M_\oplus$ (95\,\% conf.) & $<35.0\,M_\oplus$ (95\,\% conf.) & D16 \\
& $0.856_{-0.028}^{+0.028}\,M_\odot$ & & $16.5_{-2.7}^{+2.7}\,M_\oplus$ & $<12.1\,M_\oplus$ (95\,\% conf.) & $10.3_{-4.7}^{+4.7}\,M_\oplus$ &  P17 \\
\hline
$a$\,[AU] & & & $0.0811$ & $0.14120$ & $0.1879$ & C16\tablefootmark{(a)} \\
 & & & $0.08036_{-0.00088}^{+0.00088}$ & $0.1399_{-0.0015}^{+0.0015}$ & $0.1862_{-0.0020}^{+0.0020}$ & P17 \\
 & & $0.04951_{-0.00055}^{+0.00055}$ & $0.08035_{-0.00089}^{+0.00089}$ & $0.1399_{-0.0016}^{+0.0016}$ & $0.1862_{-0.0021}^{+0.0021}$ & H19\tablefootmark{(c)}  \\
 \hline
 FPP\tablefootmark{(d)} & & & $0$ & $5 \times 10^{-4}$ & $5.4 \times 10^{-6}$ & C16\tablefootmark{(a)} \\
 & & & $<1 \times 10^{-3}$ & $2.2 \times 10^{-2}$ & $<1 \times 10^{-3}$ & S16 \\
 & & & $<1 \times 10^{-4}$ & $<1 \times 10^{-4}$ & $<1 \times 10^{-4}$ & M18 \\
 & & $3.1 \times 10^{-3}$ & $5.2 \times 10^{-3}$ & $4.9 \times 10^{-3}$ & $7.2 \times 10^{-3}$ & H19 \\
\hline
\end{tabular}
\tablefoot{
           C16: \citet{2016ApJS..226....7C}; D16: \citet{2016ApJ...823..115D}; S16: \citet{2016ApJ...827...78S}; P17: \citet{2017AJ....153..142P}; M18: \citet{2018AJ....155..136M}; H19: this study.\\
           \tablefoottext{a}{Based on stellar classification ($0.92\pm0.07\,R_\odot$, $0.88\pm0.03\,M_\odot$) by \citet{2016ApJS..224....2H}.}\\
           \tablefoottext{b}{Planetary radii and 68.3\,\% confidence intervals based on our measured planet-to-star radius ratio and stellar radius estimates from P17.}\\
           \tablefoottext{c}{Semimajor axes and 68.3\,\% confidence intervals based on our measurements of the orbital periods and stellar mass estimates from P17.}\\
           \tablefoottext{d}{FPP = false positive probability. Values do not contain the ``multiplicity boost'' (reduction in FPP) but are detailed in S16.}
          }
\end{table*}

\subsection{Transit detection and characterization}

In Fig.~\ref{fig:SDE} we show the signal detection efficiency (SDE$_{\rm TLS}$) periodograms of our iterative transit search with {\tt TLS} in the K2 light curve that has been extracted with {\tt EVEREST}. Each iteration ignores the in-transit data corresponding to transits detected in previous iterations. {\tt TLS} first found planet b (top panel) because it is both the largest planet and exhibits more transits than planets c and d in the light curve. Next, {\tt TLS} found planet d (second panel), which is the second largest planet in the system. Planet c was detected in our third iteration with {\tt TLS} (third panel). In a fourth run then, we found another strong peak (SDE$_{\rm TLS}$~=~26.1) at a period of $4.34882_{-0.00075}^{+0.00069}$\,d, which we preliminary referred to as a candidate dubbed K2-32\,e. We also executed a fifth search after masking out all transits from the four planets, but we did not find any other significant signals.

Figure~\ref{fig:fold} illustrates the phase-folded light curves of K2-32\,e (top panel) and of the previously discovered planets b, c, and d from top to bottom. The shallow transit depth of about 200 parts per million (ppm) is immediately suggestive of an Earth-sized planet, given that K2-32 is a somewhat subsolar-sized K dwarf star. We note that the ordinate in the top panel, which shows the transit data of the new object, is an 18-fold zoom compared to the bottom panel, which contains the data of planets b, c, and d.

To further characterize the new planet candidate, we applied the MCMC sampler to the entire light curve. The stellar density was found to be $1.42_{-0.15}^{+0.08}$ times the solar density and the best-fit limb-darkening parameters are $q_1=0.57_{-0.18}^{+0.23}$ and $q_2=0.47_{-0.11}^{+0.16}$. Our results for K2-32\,e are shown in Table~\ref{tab:MCMC}, and our results from the MCMC sampling for the previously known planets K2-32\,b, c, and d are compared to the literature values in Table~\ref{tab:system}.

Figure~\ref{fig:architecture} is a visualization of the physical and orbital characteristics. K2-32 is indicated as a cropped orange shaded circle, and its planets are denoted by empty circles with radii to scale with the star. The vertical bars denote the uncertainties in $b$, and the transit curves are to-scale representations of the best-fit transit models shown in Fig.~\ref{fig:fold}. Orbital resonances are indicated at the bottom of this chart, with the 1:1 resonance referring to the innermost planet, K2-32\,e. Interestingly, we found that this four-planet system is close to being in a 1:2:5:7 mean motion resonance (MMR) chain. At the same time, however, the system is clearly out of this resonance compared to our measurement uncertainties.

\subsection{False-positive vetting and validation}

We used the publicly available
{\tt vespa} software \citep{2012ApJ...761....6M,2015ascl.soft03011M} to evaluate the FPP of K2-32\,e. In brief, {\tt vespa} takes the phase-folded transit light curve together with the celestial coordinates and stellar parameters to calculate the probabilities of the data being caused by non-associated blended eclipsing binaries, hierarchical triples, genuine eclipsing binaries, and non-associated stars with transiting planets. We supplied {\tt vespa} with the celestial coordinates of K2-32, $P$ and $R_{\rm p}/R_{\rm s}$ of our candidate signal (Table~\ref{tab:MCMC}), the phase-folded transit light curve (Fig~\ref{fig:fold}, top panel), the stellar parameters as determined spectroscopically by \citet{2017AJ....153..142P} (Table~\ref{tab:system}), and the 2MASS broadband photometry \citep[$J~=~10.404\,\pm\,0.024$, $H=9.993\,\pm\,0.025$, $K=9.821\,\pm\,0.019$;][]{2003yCat.2246....0C} following the ``Lessons Learned'' section in \citet{2017AJ....153..142P}. As a limiting aperture within which the transits are observed, we referred to adaptive optics and K2 light curve analyses of \citet{2016ApJ...827...78S}, who found that the transits are localized within 8 arcseconds of K2-32. {\tt vespa} returned an FPP of $3.1~\times~10^{-3}$ for K2-32\,e.

Although this number formally validates K2-32\,e as a planet already (the commonly used FFP threshold for validation is 1\,\%), it still does not consider the ``multiplicity boost'' (reduction in FPP) inherent to planet candidates in multiplanet systems.  Planet candidates in systems known to already harbor planets have a much lower FPP than single candidates \citep{2012ApJ...750..112L}. We used the values of the homogeneous K2 exoplanet survey by \citet{2016ApJS..222...14V} with $n_{\rm t}=59{,}174$ as the number of target stars (K2 campaigns 0 to 3), $n_{\rm c}=234$ as the number of K2 planet candidates, and $n_{\rm m}=26$ as the number of candidate multiplanet systems and plugged them into Eq.~(6) of \citet{2012ApJ...750..112L}. We found that the total number (the expectation value) of K2 systems like K2-32 with two or more known planets and one additional false positive ranges between 0.01, assuming a true-positive rate (or planet fidelity) of 90\,\%, and 0.05, assuming a true-positive rate of 50\,\%. These estimates increased our confidence that K2-32\,e is a true planet because both values are much lower than 1. \citet{2016ApJ...827...78S} showed that based on the statistical framework of \citet{2012ApJ...750..112L}, the multiplicity boost is at least an order of magnitude in FPP. As a consequence, a conservative estimate of the FPP for K2-32\,e including the multiplicity boost is $<\,3.1\times10^{-4}$. 

With the depth of the secondary transit ($\delta_2$) being of the order of $(R_{\rm p}/a)^2$, or 0.056\,ppm in the case of K2-32\,e, we expect this phenomenon to be invisible to K2. Thus, as an additional test for a false positive, for example, caused by an eclipsing binary, we measured $\delta_2$ in that part of the light curve where the secondary transit can be expected on a circular orbit. We detected a $8\pm8\,$ppm dip, that is to say, a $1\,\sigma$ signal that is statistically compatible with noise.

\subsection{Comparison of the {\tt BLS} and {\tt TLS} performances}

One of the most pressing questions is why K2-32\,e has not been detected in previous transit searches, for instance, those by \citet{2016ApJS..222...14V} and \citet{2016ApJS..226....7C}.

For our comparison of the transit search results obtained with {\tt BLS} and {\tt TLS}, we have to keep in mind that the definitions of the signal detection efficiency introduced by \citet{2002A&A...391..369K} for {\tt BLS} (SDE$_{\rm BLS}$) and by \citet{2019A&A...623A..39H} for {\tt TLS} (SDE$_{\rm TLS}$) are different.\footnote{See Eqs.~(5) and (6) in \citealt{2002A&A...391..369K} and Eqs.~(3) and (5) in \citealt{2019A&A...623A..39H}, respectively.} While SDE$_{\rm BLS}$ is more closely related to the mean transit depth, SDE$_{\rm TLS}$ is derived from a $\chi^2$ statistic and therefore naturally tends to produce higher values for a given transit-like signal. This principal difference between SDE$_{\rm BLS}$ and SDE$_{\rm TLS}$ must also be noted in the context of the heuristically chosen detection thresholds of typically between 8 and 9 in other studies. We have shown in \citet{2019A&A...623A..39H}, using injected transits of Earth-like planets in the light curves of Sun-like stars with white noise, that a false-positive rate of 1\,\% is achieved at virtually the same signal detection efficiency value of 7 for both {\tt BLS} (here SDE$_{\rm BLS}=7$) and {\tt TLS} (here SDE$_{\rm TLS}=7$). Although the statistics of {\tt BLS} and {\tt TLS} are different, they produced the same SDE values for this particular set of simulated data. Most important, however, the true-positive rates were $\sim~76$\,\% for {\tt BLS} and $\sim~93$\,\% for {\tt TLS}. The {\tt TERRA} algorithm uses yet another search statistic, the signal-to-noise ratio (S/N) with a detection threshold set to a nominal value of 12 ``by trial and error'' \citep{2013ApJ...770...69P}.

\citet{2016ApJS..222...14V} and \citet{2016ApJS..226....7C} both used the K2 light curve produced with the ``K2 self-flat-fielding'' ({\tt K2SFF}) pipeline, which removes most of the instrumental jitter \citep{2014PASP..126..948V}. This light curve contains a somewhat stronger noise than the light curve extracted with the {\tt EVEREST} pipeline, and we therefore need to examine whether our discovery of K2-32\,e is owing to improvements of the K2 light curve extraction and systematic detrending or due to the use of {\tt TLS} instead of {\tt BLS}. We thus repeated our analysis using the {\tt K2SFF} data and tested both {\tt TLS} and {\tt BLS}, the results of which are shown in Fig.~\ref{fig:k2sff}.

In the left column of Fig.~\ref{fig:k2sff}, we present an iterative transit search in the {\tt K2SFF} light of K2-32 with {\tt BLS}. We successively detected planets b, d, and c with very strong SDE$_{\rm BLS}$ signals (see labels) whereas planet e produced an SDE$_{\rm BLS}$ of just 8.9 in the fourth iteration, which is below the widely used detection threshold value of 9 \citep{2016ApJS..222...14V,2016ApJS..226....7C}. We verified that the addition of a smoothing filter to the SDE$_{\rm BLS}$ spectrum, similar to what is done in {\tt TLS} to the SDE$_{\rm TLS}$ spectrum, does not lift this signal above the detection threshold. None of the previous {\it K2} transit surveys by \citet{2016ApJS..222...14V}, \citet{2016ApJS..226....7C}, \citet{2018AJ....155...21P}, and \citet{2018AJ....155..136M} applied such a smoothing filter to their SDE$_{\rm BLS}$ or S/N spectra.

Moreover, the signature of K2-32\,e is only the second strongest signal, while the strongest signal near $P=0.4\,$d with an SDE$_{\rm BLS}$ of 10.5 is a false positive, as we verified. We found that this false-positive signal is produced by a combination of two characteristics of {\tt BLS}; the quantification in transit durations in units of cadences and the linear period grid typically used in {\tt BLS} applications. For example, \citet{2013ApJ...770...69P} used  ``transit durations that are integer numbers of long cadence measurements (...)'' with a grid of, for example, ``$\Delta T= [3, 5, 7, 10, 14, 18]$ long cadence measurements.'' Aliasing effects of duration and period integer multiples cause additional jitter in the {\tt BLS} spectrum, an effect that is absent in {\tt TLS} because it does not apply any binning. In addition, the period grid in {\tt TLS} is not linear \citep{2014A&A...561A.138O} because the frequency grid is linear, which intrinsically suppresses the possibility of aliasing.

We are therefore able to reproduce the results of previous searches with the traditional {\tt BLS} algorithm and confirm that K2-32\,e remains undetected in the {\tt K2SFF} light curve.

In the right column of Fig.~\ref{fig:k2sff}, we show the SDE$_{\rm TLS}$ results for an iterative transit search with {\tt TLS} in the {\tt K2SFF} data. We found planets b, d, and c with SDE$_{\rm TLS}$ values comparable to those obtained with {\tt BLS}, but most interestingly planet e also passed the detection limit with a significant SDE$_{\rm TLS}$ of 13.2. This is a fascinating real-world example of the simulation-based findings by \citet{2019A&A...623A..39H}: the strongest improvement in using {\tt TLS} instead of {\tt BLS} occurs for shallow transits or, in this case, for Earth-sized planets.

We also found that planet e can be recovered in the {\tt EVEREST} data with both {\tt BLS} (SDE$_{\rm BLS}$~=~21.3) and {\tt TLS} (SDE$_{\rm TLS}$~=~26.1). We conclude that if either the {\tt TLS} search algorithm or the {\tt EVEREST} data reduction pipeline would have been available to \citet{2016ApJS..222...14V} and \citet{2016ApJS..226....7C}, they could have found K2-32\,e. Most important, however, the combination of {\tt EVEREST} and {\tt TLS} leverages the optimal detrending and signal detection procedures.

We also examined the different methods that were used to remove longer-term stellar trends. \citet{2016ApJS..222...14V} and \citet{2016ApJS..226....7C} both used spline fits, while we used a running median by default. When we used a spline fit instead of a running median, the SDE$_{\rm TLS}$ values obtained with {\tt TLS} were virtually identical with variations of about 0.1.

\section{Discussion}

\subsection{K2-32 system}

\begin{figure*}
\centering
\includegraphics[width=.42\linewidth]{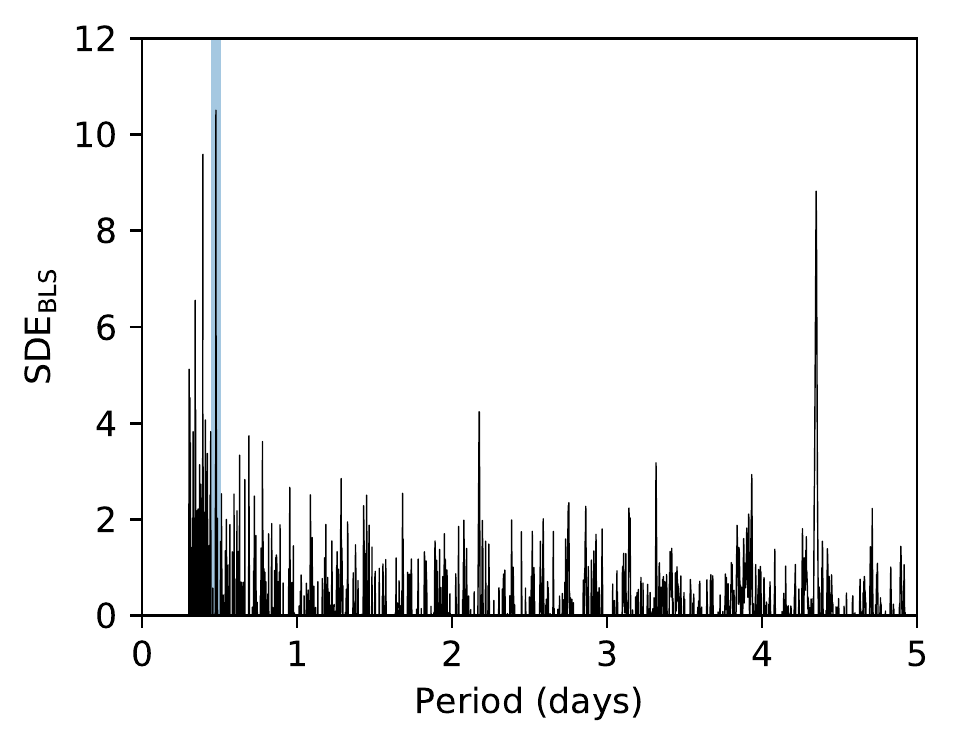}
\hspace{0.8cm}
\includegraphics[width=.42\linewidth]{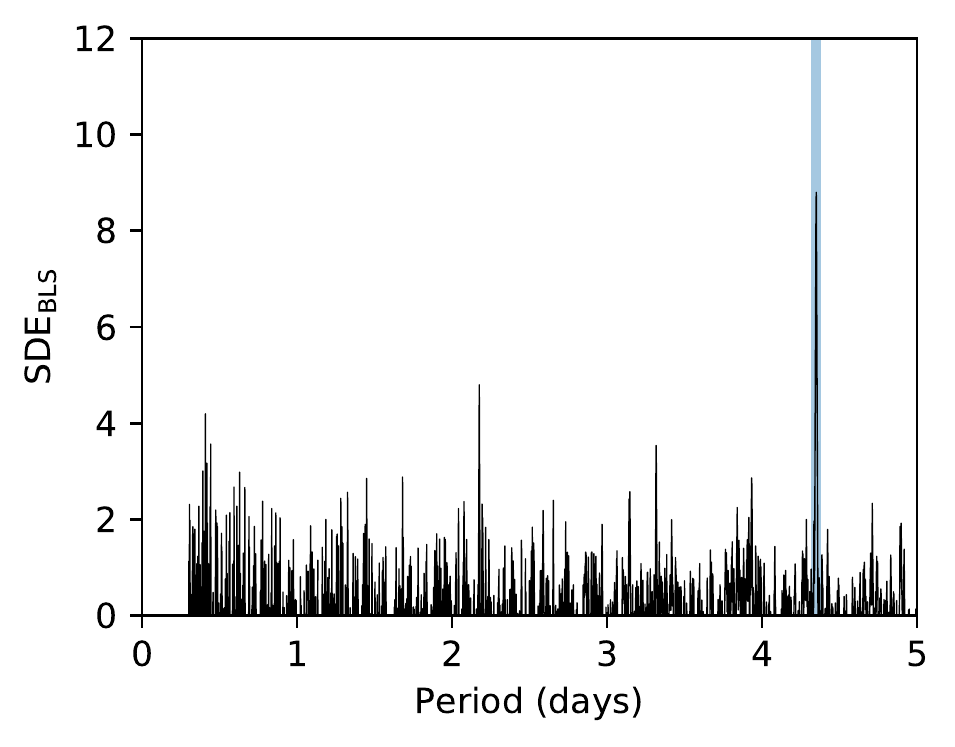}
\caption{{\it Left}: Zoom into the bottom left panel of Fig.~\ref{fig:k2sff}. The major peak near $P=0.4$\,d (blue marker) in the SDE$_{\rm BLS}$ spectrum is a false positive that is caused by an aliasing effect of the discrete cadence-based duration grid employed by most {\tt BLS} implementations, here using $10,000$ trial periods on a linear grid between 0.3\,d and 5\,d. {\it Right}: Same as in the left panel but now using more than ten times as many trial periods and a nonlinear period grid with the spacing increasing as $P^{1/3}$. Although the false positive peak near $P=0.4$\,d has disappeared, the signal of K2-32\,e near $P=4.3$\,d is still formally a false negative with SDE$_{\rm BLS}=8.9$.}
\label{fig:k2sff_bls_zoom}
\end{figure*}

For now, the mass of K2-32\,e remains unconstrained. With one Earth radius in size, a first guess for the mass of K2-32\,e is about one Earth mass. Although the densities of planets b, c, and d range between those of Saturn and Neptune, which suggests large and massive gaseous envelopes, it can safely be assumed that K2-32\,e does not carry large amounts of gas. If K2-32\,e consisted mostly of gas, its low density would imply a sub-Earth-mass planet. Such a very low-mass (and very low-gravity) planet, however, could hardly hold on to a significant gas envelope, in particular under the effects of extreme stellar irradiation. Assuming an Earth-mass ($M_\oplus$) for K2-32\,e, a nominal stellar mass of $0.856\,M_\odot$, and a semimajor axis of 0.04951\,AU (Table~\ref{tab:system}), we estimate a stellar radial velocity (RV) amplitude of $0.44\,{\rm m\,s}^{-1}$, which seems just beyond the reach of modern spectrographs \citep{2016ApJ...823..115D,2017AJ....153..142P}. Even if K2-32\,e were made up fully of iron, its mass of then roughly $1.4\,M_\oplus$ would imply an RV amplitude of just about $0.6\,{\rm m\,s}^{-1}$ and hardly surpass the RV noise (or RV jitter) of several m\,s$^{-1}$ that is typical for solar-type main-sequence stars \citep{2005PASP..117..657W}

We also studied whether transit timing variations (TTVs) could be an alternative avenue to determine the planetary masses around K2-32 \citep{2005Sci...307.1288H,2005MNRAS.359..567A}. Using our MCMC characterization of the system (Table~\ref{tab:MCMC}), a mass of $16.5\,M_\oplus$ for planet K2-32\,b \citep{2017AJ....153..142P}, and a nominal Earth mass for K2-32\,e, we find that Eq.~(8) from \citet{2012ApJ...761..122L} predicts a TTV amplitude of $\sim140$\,s or 0.04\,hr for K2-32\,e. This is probably undetectable in the available {\it Kepler} data (see top panel of Fig.~\ref{fig:fold}) but could be tested in a detailed follow-up study.

The near 1:2:5:7 MMR chain of K2-32\,e that we report is somewhat reminiscent of other MMR chains, such as the 3:4:6:8 MMR chain observed in the Kepler-223 system \citep{2016Natur.533..509M}. While the Kepler-223 system exhibits a very precise MMR tuning that has been interpreted as a footprint of planet migration, the fact that K2-32 is substantially non-synchronized with the 1:2:5:7 MMR chain suggests that additional processes have been at work. The origin of these deviations from precise commensurability might be in long-term star-planet tidal interaction that was only overruled by the MMR-creating effects of the protoplanetary disk in the very early stages of the system \citep{2011CeMDA.111...83P,2018arXiv180606601H}. K2-32 joins the family of planetary systems from the {\it Kepler} mission that are just wide of exact commensurability \citep{2013ApJ...774...52L}, although K2-32 represents a particular case with four rather than just two near-resonant planets. \citet{2011ApJS..197....8L} defined a new variable

\begin{equation}
\zeta_1=3\left( \frac{1}{\mathcal{P}-1} - {\rm Round}{\Big (} \frac{1}{\mathcal{P}-1}{\Big )}\right)
\end{equation}

\noindent
as a measure of the difference between an observed first-order MMR period ratio, where $\mathcal{P}=P_{\rm o}/P_{\rm i}$ is the orbital period ratio between the outer and the inner planet. These authors found statistically significant deviations of the observed $\zeta_1$ distribution from a random period ratio distribution for about 100 transiting exoplanets from {\it Kepler} known at the time. Their results showed that for a randomly drawn pair of {\it Kepler} planets, $\zeta$ is most likely to range between -0.1 and -0.2. Taking the median orbital periods from our MCMC sampling for K2-32\,e and b (Table~\ref{tab:MCMC}), which are very close to the 2:1 MMR resonance ($P_{\rm o}/P_{\rm i}=2.0676$), we find $\zeta_1=-0.19$. {Although this result cannot be used as a means of verification for K2-32\,e, it is in very good agreement with a large sample of exoplanets from {\it Kepler} \citep{2011ApJS..197....8L} and, therefore serves as an indirect piece of evidence in favor of the planetary nature of K2-32\,e.

\subsection{Aliasing effects in {\tt BLS}}

Figure~\ref{fig:k2sff} shows the results we obtained with the {\tt BLS} reference implementation in {\tt astropy}. Nevertheless, the question arises if the occurrence of a false positive near $P=0.4$\,d could be avoided with different grids for the trial periods and trial durations. We therefore varied the resolution and spacing of both the duration and the period grids for {\tt BLS} to determine whether we could remove this alias peak. The results are illustrated in Fig.~\ref{fig:k2sff_bls_zoom} and are discussed in the following.

First, the grid of trial transit durations used in {\tt BLS} must be in multiples of a constant cadence. As an intrinsic property of {\tt BLS}, any given cadence is tested for one of two test flux values that relate to either the out-of-transit flux or to the in-transit flux \citep{2002A&A...391..369K}. As a consequence, when {\tt BLS} is applied with a period grid of constant spacing, short trial periods may happen to be integer multiples of the trial duration. For example, a trial period of 0.416\,d is sufficiently close to the $20$-fold multiple of a 30\,min trial transit duration (a single long-cadence exposure of a {\it K2} light curve), and a 10-fold multiple of a 60\,min transit duration (worth two cadences), etc. As the transit duration grows with the orbital period to the power of one-third \citep[Eq.~(10) in][]{2019A&A...623A..39H}, everything else being equal, aliasing becomes less of an issue for longer trial periods. Nevertheless, aliasing cannot be avoided completely by using an infinitely fine period grid. Instead, it can be avoided by using a nonlinear period grid as employed by {\tt TLS}.

Second, the grid of trial durations tested with {\tt BLS} is usually the same for all trial periods, see, for example, the {\tt BLS} implementation in {\tt astropy}. Although users can define a maximum trial duration of, for example, $\sim0.21\,$days (or 10 cadences), which is sufficiently long for most long-period planets in {\it K2}, this trial duration is as much as half of an orbital period for the most short-period planets that are physically plausible. Again, clustering of data can generate aliases such as those visible near $P=0.3$\,d in Fig.~\ref{fig:k2sff_bls_zoom} (left panel) and eventually lead to false-positive detections. In this example, the most reliable way for us to remove the alias peaks was to use a hyperfine nonlinear grid of $>100,000$ trial periods in combination with a duration grid that uses a maximum trial duration shorter than half of the shortest period. Nevertheless, and maybe most important, the signal of K2-32\,e near $P=4.3$\,d remained a false negative for the {\tt BLS} search.

In practice, it could be possible to develop more appropriate trial period and trial duration grids for {\tt BLS} similar to the grid that is implemented in {\tt TLS}, but this is beyond the scope of this paper and the disadvantage of the suboptimal detector shape (box versus limb-darkened transit) would remain.

\section{Conclusions}

We determined with high significance (SDE$_{\rm TLS}$~=~26.1) a fourth sequence of periodic transits in the light curve of K2-32. Our MCMC simulations suggest that this signal has a period of $4.34882_{-0.00075}^{+0.00069}$\,d and the transiting planet has a radius of ${1.01}_{-0.09}^{+0.10}\,R_\oplus$, making it one of the smallest planets found with K2 so far. Our false-positive vetting of K2-32\,e as an individual transiting object yields ${\rm FPP}=3.1\times10^{-3}$. Factoring in the planetary multiplicity around K2-32, we find ${\rm FPP}<3.1\times10^{-4}$. This formally validates K2-32\,e as a planet.

This new planet reveals a near 1:2:5:7 MMR chain of now four planets around K2-32. While being very close to this MMR chain, however, the planets are in fact just wide of exact commensurability, thereby joining a growing family of this type of systems from the {\it Kepler} mission. We find that the offset from exact commensurability between the new planet and the previously know K2-32\,b is in very good agreement with the offsets found in other multiplanet transiting systems, adding more evidence in favor of the planetary nature of K2-32\,e. K2-32 also joins the list of {\it K2} systems with four or more transiting planet candidates, about a dozen of which are known as of today. For now, K2-138 is the only system with five \citep{2018AJ....155...57C} or potentially even six \citep{2019AAS...23316407H} transiting planets, all of which are in the super-Earth to sub-Neptune radius regime.

Our discovery confirms that {\tt TLS} can find sub-Earth-sized planets that have previously been missed with search algorithms looking for box-like transit signals such as {\tt BLS} \citep{2002A&A...391..369K}. We verified that there are two reasons why previous searches have missed K2-32\,e. First, we used the K2 light curve of K2-32 that was subject to the highly efficient removal of systematic effects with {\tt EVEREST} \citep{2016AJ....152..100L}, while previous searches used the light curve detrended with {\tt K2SFF} that has slightly less favorable noise properties. Second, {\tt TLS} is intrinsically more efficient in finding shallow transits because the search function is a transit and not a box \citep{2019A&A...623A..39H}. Interestingly, we find that K2-32\,e could have been detected in the {\tt K2SFF} light curve with {\tt TLS}.

\begin{acknowledgements}
The authors thank the referee for a helpful report. This research has made use of the NASA Exoplanet Archive, which is operated by the California Institute of Technology, under contract with the National Aeronautics and Space Administration under the Exoplanet Exploration Program. This work made use of NASA's ADS Bibliographic Services. RH is supported by the German space agency (Deutsches Zentrum f\"ur Luft- und Raumfahrt) under PLATO Data Center grant 50OO1501. RH also wishes to thank the Hans and Clara Lenze Foundation in Aerzen for financial support. KR is a member of the International Max Planck Research School for Solar System Science at the University of G\"ottingen. KR performed the MCMC analysis of the light curve.
\end{acknowledgements}

\bibliographystyle{aa}
\bibliography{aa}
\end{document}